\documentclass[10pt,conference]{IEEEtran}

\IEEEoverridecommandlockouts
\usepackage{booktabs}
\usepackage{url}
\usepackage{hyperref}
\usepackage{tabularx}
\usepackage{cite}
\usepackage{amsmath,amssymb,amsfonts}
\usepackage{algorithmic}
\usepackage{graphicx}
\usepackage{textcomp}
\usepackage{xcolor}
\def\BibTeX{{\rm B\kern-.05em{\sc i\kern-.025em b}\kern-.08em
    T\kern-.1667em\lower.7ex\hbox{E}\kern-.125emX}}
\begin{document}

\title{Ensuring Robustness in ML-enabled Software Systems: A User Survey}


\author{\IEEEauthorblockN{Hala Abdelkader}
    \IEEEauthorblockA{\textit{Applied Artificial Intelligence Institute} \\
    \textit{Deakin University}\\
    VIC, Australia \\
    habdelkader@deakin.edu.au}
    \and
    \IEEEauthorblockN{Mohamed Abdelrazek}
    \IEEEauthorblockA{\textit{Applied Artificial Intelligence Institute} \\
    \textit{Deakin University}\\
    VIC, Australia \\
    mohamed.abdelrazek@deakin.edu.au}
    \and
    \IEEEauthorblockN{Priya Rani}
    \IEEEauthorblockA{\textit{STEM - School of Engineering} \\
    \textit{RMIT University}\\
    VIC, Australia \\
    priya.rani@rmit.edu.au}
    \and
    \IEEEauthorblockN{Rajesh Vasa}
    \IEEEauthorblockA{\textit{Applied Artificial Intelligence Institute} \\
    \textit{Deakin University}\\
    VIC, Australia \\
    rajesh.vasa@deakin.edu.au}
    \and
    \IEEEauthorblockN{Jean-Guy Schneider}
    \IEEEauthorblockA{\textit{Department of Software Systems \& Cybersecurity} \\
    \textit{Monash University}\\
    VIC, Australia \\
    jean-guy.schneider@monash.edu}
   
}

\maketitle

\begin{abstract}
Ensuring robustness in ML-enabled software systems requires addressing critical challenges, such as silent failures, out-of-distribution (OOD) data, and adversarial attacks. Traditional software engineering practices, which rely on predefined logic, are insufficient for ML components that depend on data and probabilistic decision-making. To address these challenges, we propose the ML-On-Rails protocol, a unified framework designed to enhance the robustness and trustworthiness of ML-enabled systems in production. This protocol integrates key safeguards such as OOD detection, adversarial attack detection, input validation, and explainability. It also includes a model-to-software communication framework using HTTP status codes to enhance transparency in reporting model outcomes and errors. To align our approach with real-world challenges, we conducted a practitioner survey, which revealed major robustness issues, gaps in current solutions, and highlighted how a standardised protocol such as ML-On-Rails can improve system robustness. Our findings highlight the need for more support and resources for engineers working with ML systems. Finally, we outline future directions for refining the proposed protocol, leveraging insights from the survey and real-world applications to continually enhance its effectiveness.
\end{abstract}


\section{Introduction}
The transition of machine learning (ML) models from prototyping to production environments presents significant challenges regarding safety, security, and transparency~\cite{arnold2019factsheets, hendrycks2021unsolved}. These challenges directly impact the robustness and trustworthiness of ML-enabled software systems~\cite{scher2022robustness, hendrycks2021unsolved, foidl2019risk, sutrop2019should, arnold2019factsheets, khomh2018software}. One of the most critical issues is the silent failure of ML models, where incorrect results are generated without any visible indication of error~\cite{abdelkader2024ml, abdelkader2023robustness, cummaudo2019, rabanser2019}. Similarly, OOD data, where production data deviates from the training data distribution, leads to performance degradation and inaccurate predictions~\cite{sun2022dice, sun2021react, yang2021generalized}. Additionally, adversarial attacks pose a critical security threat, where malicious inputs are crafted to deceive the model into generating faulty outputs~\cite{macas2023adversarial, goodfellow2014, kurakin2018adversarial, huang2017adversarial}.

Addressing these concerns requires robust safeguards that can operate effectively in production environments~\cite{abdelkader2024ml, abdelkader2023robustness}. However, traditional software engineering practices for ensuring robustness fall short when applied to ML components due to the fundamental differences between traditional systems and ML models~\cite{amershi2019software, kumeno2019software}. Traditional software relies on predefined logic and explicit rules to produce deterministic outputs, while ML models use data-driven techniques to optimise their internal logic through iterative, data-driven learning~\cite{amershi2019software, kumeno2019software}. This fundamental difference results in a `black box' effect, where learned patterns replace explicit rules, resulting in less predictable behavior~\cite{amershi2019software, kumeno2019software, abdelkader2024ml}. Also, the development of ML models involves three distinct stages: data, model, and management. These stages are unique to ML systems and not the norm for traditional software, which further complicates efforts to ensure robustness~\cite{abdelkader2024ml, abdelkader2023robustness, amershi2019software, kumeno2019software}.


While various solutions have been proposed to address individual robustness challenges, ML models are often deployed without integrated robustness measures~\cite{abdelkader2024ml}. This leaves software engineers responsible for maintaining model robustness without a unified framework to enforce ML model specifications and robustness in production settings~\cite{abdelkader2024ml}. Additionally, the absence of standardised terminology for communicating ML model behavior creates barriers to effective collaboration between data scientists and engineers~\cite{abdelkader2024ml, cummaudo2019}.

Therefore, we propose the ML-On-Rails protocol, a comprehensive solution designed to enhance the robustness of ML models in production environments. The protocol integrates:

\noindent\textbf{ML safeguards:} A set of mechanisms, including OOD detection, adversarial attack detection, explainability, and input validation, designed to address the robustness challenges of ML-enabled software systems.

\noindent\textbf{Model-2-Software communication framework:} A structured communication framework that uses HTTP status codes to systematically report model outcomes and errors, ensuring transparency in ML-enabled software systems.

To validate the protocol, we conducted a user survey to capture real-world challenges engineers encounter when working with ML components in production environments. The primary objectives of this survey are:

\begin{itemize}
    \item To pinpoint areas where software engineers face difficulties and require additional support or resources when working with ML-enabled systems in production.
    \item To evaluate the effectiveness of the proposed ML-On-Rails protocol in addressing the identified challenges.
    \item To refine the protocol to better meet the practical needs of engineers and contribute to the development of more robust ML-enabled systems.
\end{itemize}

\smallbreak
Our survey aims to answer the following research questions: 

\noindent\textbf{RQ1:} What are the most common robustness challenges encountered by ML-enabled systems?

\noindent\textbf{RQ2:} What limitations or gaps exist in current solutions for addressing these robustness challenges in ML systems?

\noindent\textbf{RQ3:} How can a standardised protocol enhance the robustness, trustworthiness, and usability of ML models in production? 

\noindent\textbf{RQ4:} What key information do software engineers need about ML model behavior?
\smallbreak

The rest of the paper is organised as follows: Section~\ref{sec:protocol} introduces the ML-On-Rails protocol, outlining its design and safeguards components. Section~\ref{sec:survey} covers the practitioner survey, participant selection criteria, and key findings. Section~\ref{sec:discussion} analyses the results, suggests improvements, and discusses future research. The paper concludes with a summary in Section~\ref{sec:conclusion}.

\begin{figure}[t]
\centering
    \includegraphics[width=\linewidth]{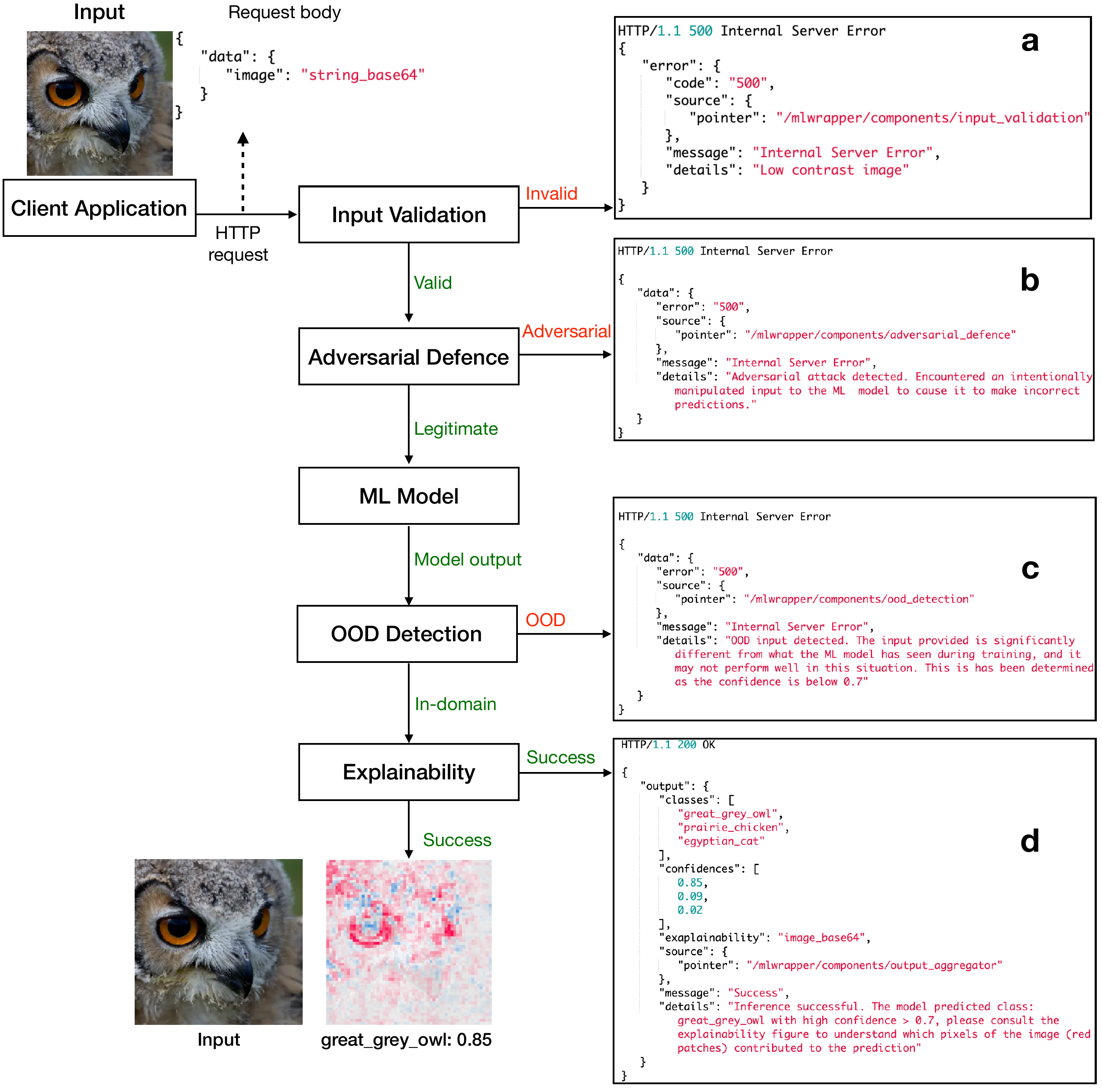}
    \caption{An illustrative example of the proposed ML-On-Rails protocol that shows the sequence of operations of a given input and possible outputs}
    \label{fig:user_survey_protocol}
\end{figure}

\section{ML-On-Rails protocol}
\label{sec:protocol}
The ML-On-Rails protocol is a unified framework designed to enhance the robustness of ML-enabled systems in production. This section provides an overview of its key components.

\subsection{Input validation Safeguard}
Input validation is crucial in machine learning to ensure data quality and compatibility with model requirements. In computer vision, this involves verifying image formats, checking dimensions, and evaluating contrast and resolution to prevent low-quality inputs from degrading model performance. These steps help maintain high data standards, ensuring reliable and accurate model predictions.

\subsection{Model Explainability Safeguard}

Model explainability contributes to building trust in ML-enabled systems~\cite{molnar2020interpretable, ribeiro2016should}. In production, understanding the rationale behind model predictions is crucial for debugging, enhancing model performance, and addressing potential biases~\cite{molnar2020interpretable}. SHAP (Shapley Additive exPlanations) is a widely used explainability method that reveals how individual features impact predictions~\cite{lundberg2017unified}. Derived from cooperative game theory, SHAP Values create a comprehensive framework for assessing feature importance in model predictions, applicable to any ML model (model agnostic). Shapley Values fairly distribute the model prediction among the features based on their contribution~\cite{molnar2020interpretable}. SHAP is a powerful visualisation tool that shows how each feature contributes to ML model predictions.
SHAP values reveal the impact of each feature on model predictions for individual instances.
In the proposed ML-On-Rails protocol, SHAP can be effectively integrated as an explainability tool to accompany model predictions.

\subsection{Out-of-Distribution Data Detection Safeguard}

OOD data refers to data that follows a different distribution from the training data used to develop an ML model~\cite{sun2022dice, sun2021react, yang2021generalized, liang2018, hendrycks2017}. These data lie outside the distribution the model was trained on and often represent scenarios or conditions the model did not encounter in the training phase~\cite{sun2022dice, sun2021react, yang2021generalized, liang2018, hendrycks2017}. This inconsistency between training and deployment distributions is likely to cause poor model performance and degrade the accuracy of ML-enabled software systems and services~\cite{koh2021wilds, yang2021generalized, sun2022dice, sun2021react}. Maximum Softmax Probability (MSP)~\cite{hendrycks2017} is the baseline for OOD detection techniques where the softmax probabilities are used to classify OOD inputs. It relies on the fact that the ML model would produce highly confident predictions with in-domain inputs compared to OOD inputs. The maximum softmax class probabilities of the model for an input test example are evaluated against a threshold to determine whether it is an OOD input.

\subsection{Adversarial Attacks Detection Safeguard}

Adversarial attacks are generated by intentionally introducing small, unnoticeable perturbations to input data that cause the ML models to make incorrect predictions~\cite{goodfellow2014, kurakin2018adversarial, huang2017adversarial}. These perturbations are often imperceptible to humans but can impact the model's performance~\cite{goodfellow2014, kurakin2018adversarial, huang2017adversarial}. In our proposed protocol, we employ adversarial training as a defense method. The adversarial training approach involves enriching the training set with perturbed data to improve model robustness~\cite{goodfellow2014, szegedy2014intriguing, lyu2015unified}. This approach enables the model to learn from both original and perturbed inputs, while an auxiliary model is trained specifically to detect adversarial attacks~\cite{chakraborty2018adversarial}. 

\subsection{Model-2-Software Communication Framework}
Silent failures in production ML models can be addressed using error codes, which help developers identify and resolve performance issues. These codes offer insights into specific failures, improving model robustness. HTTP status codes can be used to indicate both successes and errors in model decisions. For example, HTTP code 200 signals successful execution, including classification results and decision rationale, while HTTP code 500 flags issues like invalid input or adversarial attacks. This system improves transparency and provides a clear starting point for diagnostics, helping developers understand and fix errors.

\section{Practitioner survey}
\label{sec:survey}
The survey consisted of four sections:

\begin{enumerate}
    \item Background and demographics: collected participants' demographic details, including their roles, experience with ML-enabled systems, and the number of deployed systems they worked on. It also explored the sources of ML components used and the types of applications they typically work on.
    \item Challenges with ML models: identified challenges software engineers face with ML models, including understanding jargon, integrating models into systems, monitoring production environments, and handling issues like model performance and behavior deviations.
    \item Proposed ML-On-Rails protocol: participants provided feedback on the essential features for ensuring the robustness of ML components in production, as well as additional considerations for the protocol’s design.
    \item ML-On-Rails protocol evaluation: given the example illustrated in Fig.~\ref{fig:user_survey_protocol}, participants evaluated the protocol’s effectiveness in addressing issues like input validation, OOD detection, and adversarial attacks, and gave feedback on ease of use, clarity of messages, and potential improvements.

\end{enumerate}


\subsection{Recruitment of Participants}

The target population included professionals with experience building or managing AI/ML-enabled systems, such as software engineers, data scientists, and researchers. We used convenience sampling to select participants based on their accessibility and availability, aiming to recruit at least 15 participants. We utilised contact networks on platforms like LinkedIn, Reddit, and software engineering mailing lists. We also employed snowball sampling by encouraging participants to share the survey with colleagues and contacts who meet the criteria. The survey was anonymous, with no incentives, and took approximately 25 minutes to complete. Out of the 85 participants who provided consent, only 33 fully completed the survey. Partial responses were excluded as outlined in the Plain Language Statement (PLS). After applying the selection criteria in Table~\ref{tab:inclusion_exclusion_criteria}, 30 participants with relevant experience in ML-enabled software systems were included.

\begin{table}[tb]
  \centering
  \renewcommand{\arraystretch}{0.95}
  \caption{Inclusion and exclusion criteria for the survey responses}
  \label{tab:inclusion_exclusion_criteria}
  \begin{tabular}{p{0.45\linewidth} p{0.45\linewidth}}
    \hline
    \textbf{Inclusion Criteria} & \textbf{Exclusion Criteria} \\ 
    \hline
    Consent provided & Consent not provided \\
    Survey completed & Partial responses \\ 
    Experience with ML-enabled Software Systems & No experience with ML-enabled Software Systems \\
    \hline
  \end{tabular}
\end{table}

\subsection{Results}

\begin{figure}[t]
  \centering
  \includegraphics[width=\linewidth]{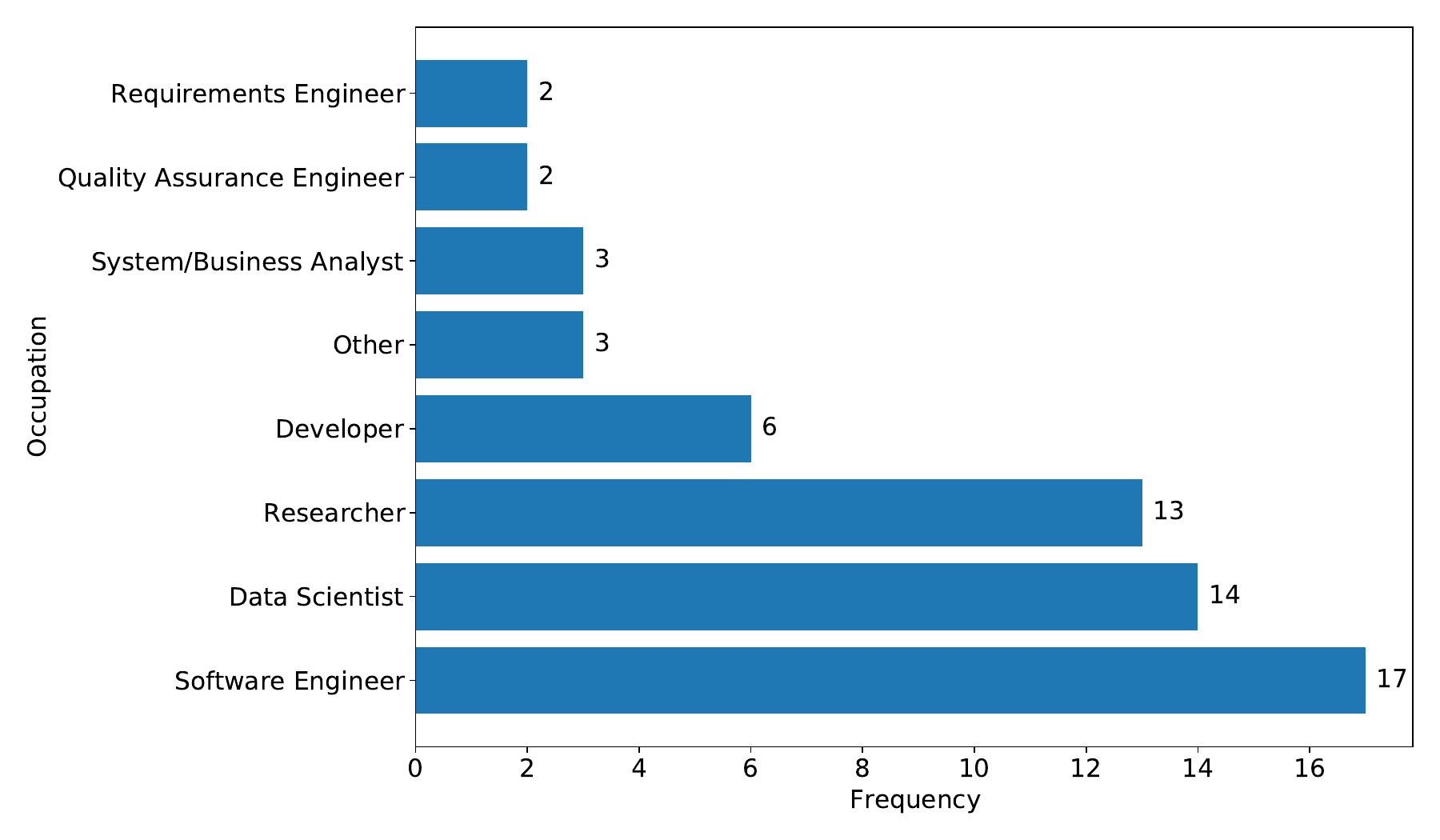}
  \caption{Participant roles within the organization}
   \label{fig:occupation_distribution}
\end{figure}

The majority of respondents are Software Engineers, with a total of 17. This is followed by 14 Data Scientists and 13 Researchers. Six participants identified themselves as Developers, followed by three as System/Business Analysts, two as Quality Assurance Engineers, and two as Requirements Engineers. This distribution suggests a diverse range of expertise among the survey participants, including engineering, research, and data science as shown in Fig.~\ref{fig:occupation_distribution}. 

We asked the participants about their expertise in building ML-enabled systems. The most common experience level is four years, reported by 10 participants (33.3\%). Nine participants have two years of experience, followed by seven participants with seven years of experience. A smaller number of participants have more years of experience: one participant has thirteen years, another has sixteen years, and one has eighteen years of experience. This distribution indicates that the survey participants cover a wide range of experience levels, with most being early to mid-career professionals. Although there is representation across various stages of expertise, the majority of participants have less than ten years of experience, suggesting that the insights gathered primarily reflect the perspectives of those earlier in their careers.

We inquired about the number of ML-enabled systems the participants have worked on and deployed in production. Eleven participants have worked on three or more deployments, while ten participants have deployed at least one system. Notably, six participants reported working on five or more deployments. However, three participants have not deployed any ML systems in production; these individuals are researchers with 9 and 4 years of experience and a quality assurance engineer with 18 years of experience. This distribution highlights that a significant portion of the participants have substantial experience with ML-enabled software systems in production, with 17 participants having worked on at least three systems. Conversely, a smaller number of participants, 3 in total, have not had experience with deployed systems. This suggests that most participants are reasonably experienced with deploying ML systems, which provides valuable insights into practical challenges and considerations in production.

\smallbreak
\noindent\textbf{ML Component Sources and Application Domains:}

We asked participants about the sources of ML components they have used, allowing them to choose multiple options. The results indicate that custom-developed ML models were the most popular choice (35.4\%), selected by 23 participants. Another 14 participants chose pre-trained off-the-shelf models, valuing the balance between convenience and effectiveness. Additionally, 17 participants used customised off-the-shelf models, which involved further tuning pre-trained models to fit specific project requirements. Finally, 11 participants relied on third-party models accessed via APIs, reflecting a preference for integrating ML capabilities without direct management.

\begin{figure}[t]
  \centering
  \includegraphics[width=\linewidth]{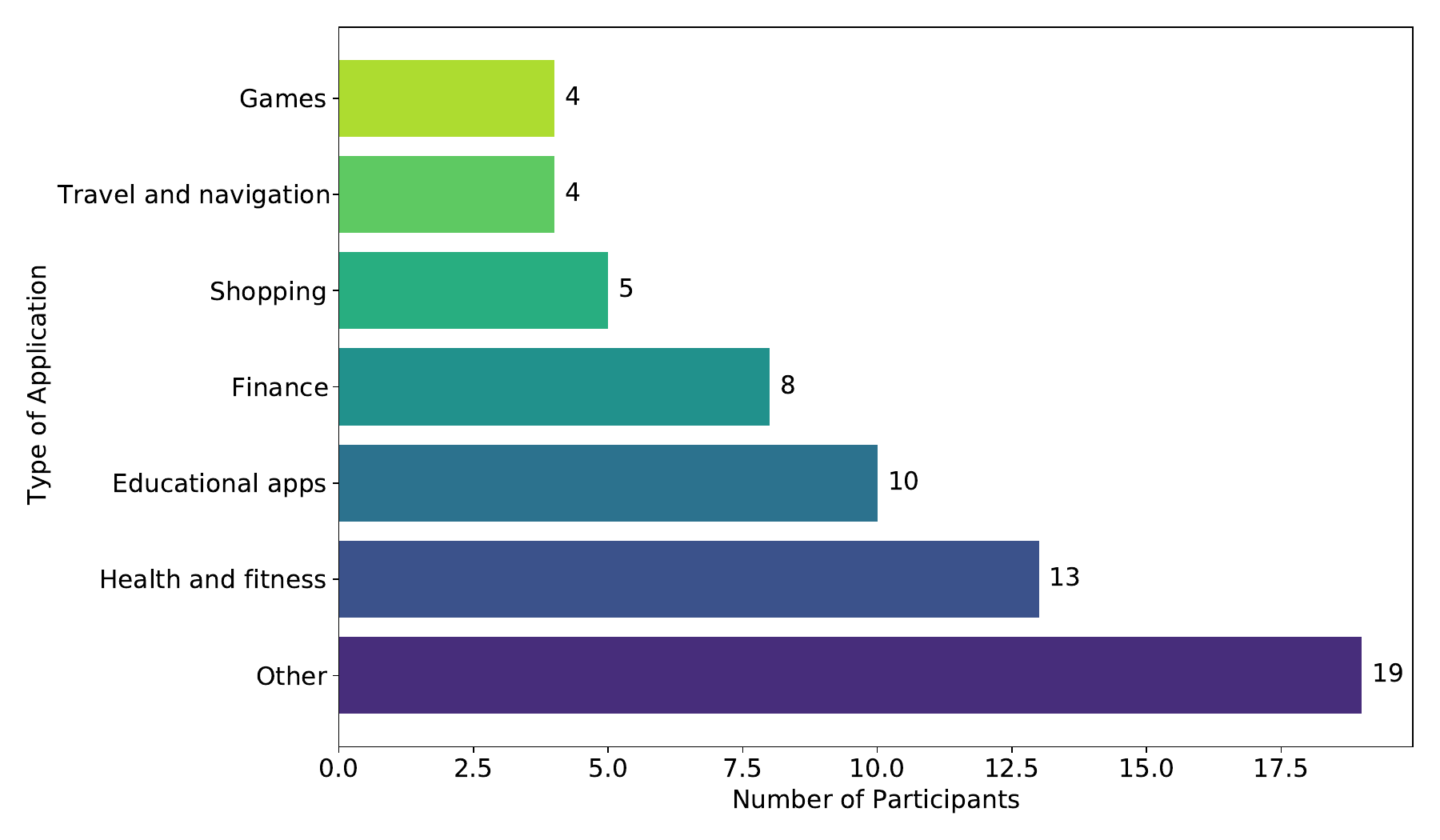}
  \caption{Distribution of application types among participants}
   \label{fig:application}
\end{figure}

We further asked the participants about the types of applications they commonly work on. The responses to this question reveal an interesting distribution of focus areas, as illustrated in Fig.~\ref{fig:application}. Thirteen participants worked on health and fitness applications, while ten participants focused on educational apps. Finance applications are the focus of eight participants. Shopping applications are handled by five participants. Travel and navigation, as well as games, each have four participants working in these areas. Notably, nineteen participants selected ``Other" indicating a broad range of application types that may not fit into the predefined categories. This suggests a diverse array of applications not fully captured by the given options.

\smallbreak
\noindent\textbf{Challenges Software Engineers Face with ML Models:}

\begin{figure}[t]
 \centering
  \includegraphics[width=\linewidth]{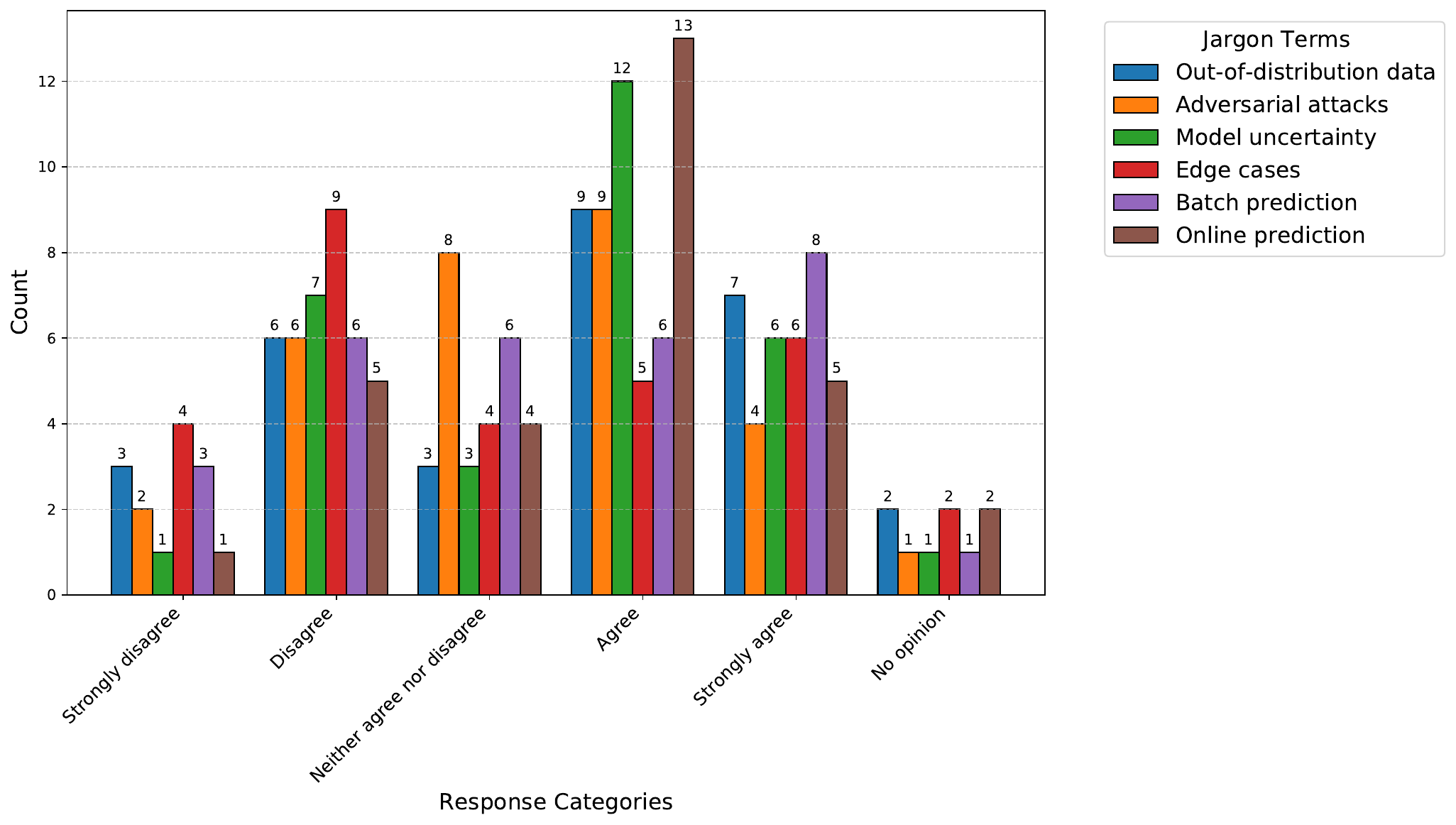}
  \caption{Responses on challenges with machine learning terminology}
   \label{fig:jargon}
\end{figure}

We asked the participants about the challenges they face in understanding and communicating specific machine-learning jargon. The terms considered in the survey are: ``Out-of-distribution data", ``Adversarial attacks", ``Model uncertainty", ``Edge cases", ``Batch prediction", and ``Online prediction". The responses are categorised from ``Strongly disagree" to ``Strongly agree", offering insights into the varying levels of familiarity and difficulty associated with these terms among the participants, as illustrated in Fig.~\ref{fig:jargon}.

Model uncertainty, Out-of-distribution data, and Online prediction are identified as the most challenging terms, as indicated by the highest number of ``Agree" and ``Strongly agree" responses. Adversarial attacks and Batch prediction are considered moderately challenging. Finally, Edge cases term has mixed responses, with a notable portion of respondents indicating that they do not find it particularly challenging. Overall, fundamental concepts in machine learning, such as Model uncertainty and Out-of-distribution data, tend to be more challenging. These insights can inform communication strategies for software engineers by focusing on the most challenging terms and providing explanations where necessary.

\smallbreak
We also asked participants to identify the key challenges of integrating ML models into software systems and select their top five. The distribution of integration challenges is shown in Table~\ref{tab:integration_challenges}, and the most critical challenges reported were: 
\begin{itemize}
    \item \textbf{Lack of explainability of model outputs:} 20 participants
    \item \textbf{Model versioning:} 19 participants
    \item \textbf{Unrealistic expectations about model capabilities:} 15 participants
\end{itemize}

A recurring theme in the responses was insufficient documentation, particularly regarding how the model works, how it was trained, what datasets were used, and how to use it. These responses underscore the need for comprehensive documentation. Improving these aspects is crucial for facilitating integration and ensuring ongoing maintenance. Participants also noted gaps in available testing and debugging tools and highlighted collaboration challenges due to mixed team dynamics. This suggests a strong need for better tools and processes to address ML-specific issues and facilitate improved collaboration. Finally, challenges related to evaluation metrics and complex entanglements reflect the need for robust methodologies and management practices to deal with ML model interactions and performance assessment. These challenges are key barriers to the successful integration and effective use of ML models. 


\begin{table}[t]
    \centering
    \renewcommand{\arraystretch}{0.95}
    \caption{Challenges in integrating ML models into software systems}
    \label{tab:integration_challenges}
    \begin{tabularx}{\linewidth}{X|c}
    \toprule
    \textbf{Integration Challenge} & \textbf{Frequency} \\ \midrule
    Lack of explainability of model outputs & 20 \\ \hline
    Model versioning & 19 \\ \hline
    Unrealistic expectations about model capabilities & 15 \\ \hline
    Insufficient documentation on how the model works & 14 \\ \hline
    Insufficient documentation on how the model was trained and what datasets were used & 14 \\ \hline
    Insufficient documentation on how to use the model & 13 \\ \hline
    Current software testing and debugging tools not applicable to ML components & 10 \\ \hline
    Collaboration challenges due to mixed team dynamics, skills, and tools used & 10 \\ \hline
    Evaluation metrics & 10 \\ \hline
    Complex entanglement caused by dependencies between multiple ML models in a software system & 10 \\ \hline
    How to test ML model components in software systems & 9 \\ \hline
    ML packages do not check for input validity (interfaces) & 6 \\ \bottomrule
    \end{tabularx}

\end{table}


\begin{table}[t]
    \centering
    \renewcommand{\arraystretch}{0.95}
    \caption{Challenges with ML models in production software systems}
    \label{tab:production_challenges}
    \begin{tabularx}{\linewidth}{X|c}
    \toprule
    \textbf{Production Challenge} & \textbf{Frequency} \\ \midrule
    Silent Failure & 19 \\ \hline
    Out-of-distribution data (Training-serving skew) & 17 \\ \hline
    Lack of transparency and explainability & 17 \\ \hline
    Output is unpredictable & 15 \\ \hline
    Model uncertainty & 12 \\ \hline
    Adversarial attacks (ML models can easily be fooled) & 9 \\ \hline
    Lack of automation (of the training \& deployment pipeline) & 9 \\ \hline
    Monitoring quality attributes such as fairness and bias & 9 \\ \hline
    Addressing changing customer needs & 8 \\ \hline
    Evolution of ML services & 8 \\ \hline
    Lack of human-driven evaluation (human in the loop) & 7 \\ \hline
    Exception handling & 6 \\ \hline
    Concerns about accountability and ethics of ML & 5 \\ \hline
    Input validation & 4 \\ \hline
    Troubleshooting system failures & 4 \\ \hline
    No universally accepted ethical system for ML & 1 \\ 
    \bottomrule
\end{tabularx}

\end{table}
\smallbreak
We asked participants to identify the common challenges they face with ML models deployed in production software systems and select their top five. The distribution of integration challenges is shown in Table~\ref{tab:production_challenges}, and the responses revealed several key trends:
\begin{itemize}
    \item \textbf{Silent failure:} The most reported challenge (19 participants) is silent failures, where ML models make incorrect predictions without raising any errors or warnings. This indicates that communicating issues related to model malfunctions is the top concern for many participants.
    \item \textbf{OOD data and lack of transparency:} Both OOD data (17 participants) and lack of transparency and explainability (17 participants) are significant challenges. The high frequency of these responses suggests that training-serving skew and difficulties in understanding model behavior are substantial challenges in deployments.
    \item \textbf{Unpredictable outputs and model uncertainty:} Unpredictable outputs (15 participants) and model uncertainty (12 participants) are also common concerns.
    \item \textbf{Adversarial attacks and monitoring concerns:} Although less frequent, adversarial attacks (9 participants), lack of automation (9 participants), and challenges in monitoring fairness and bias (9 participants) are still existing concerns. 
\end{itemize}

Overall, the primary concerns focus on the safety, transparency, and behavior of ML models in production environments. Participants identified key challenges such as silent failures, OOD data, and lack of transparency and explainability, highlighting the need for better monitoring, detection, and interpretability tools. Addressing these issues is crucial for ensuring the robustness of ML-enabled software systems.

\smallbreak
\noindent\textbf{Identifying ML Model Issues in Production:}

\begin{figure}[t]
  \centering
  \includegraphics[width=\linewidth]{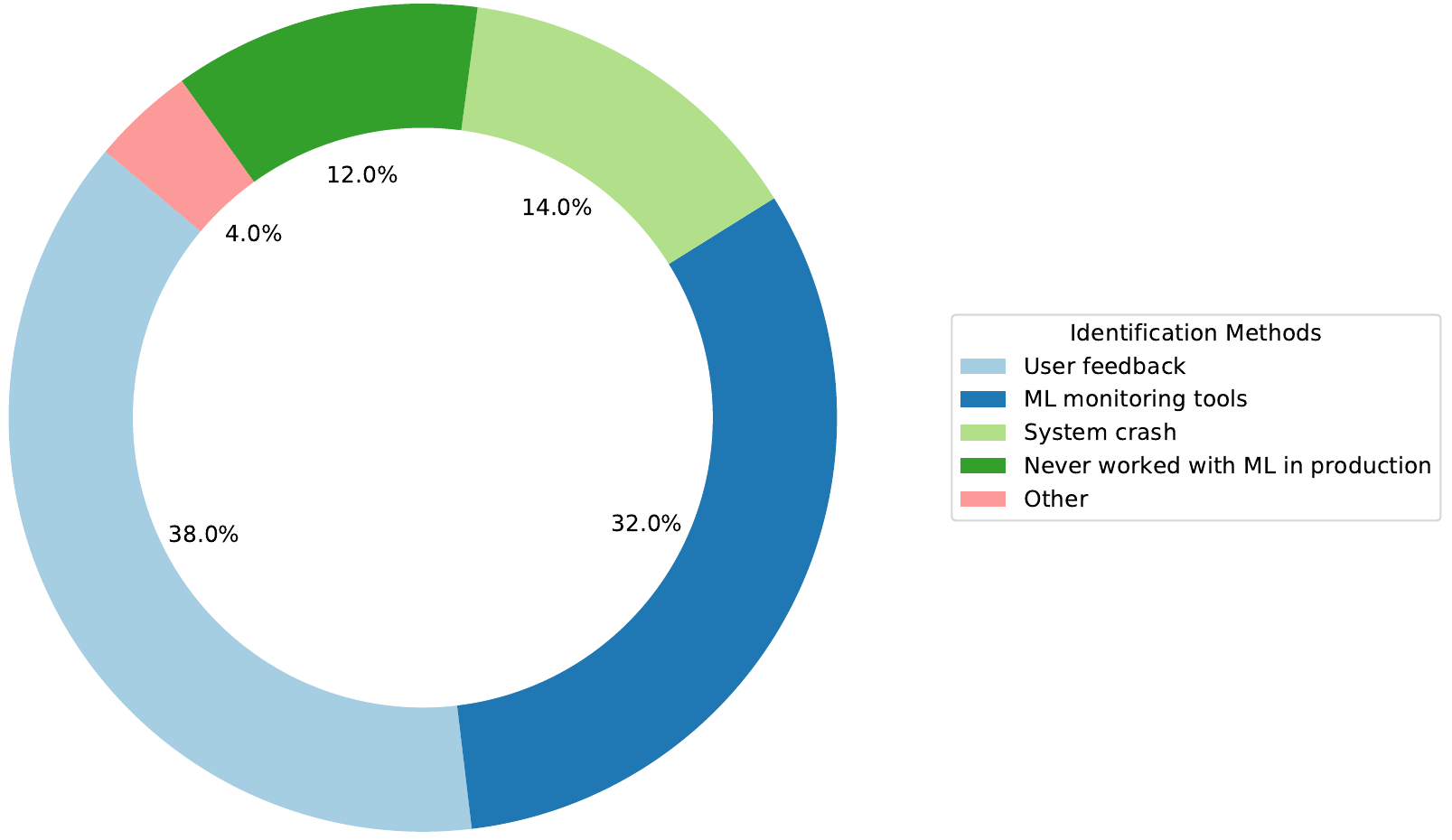}
  \caption{Process for identifying ML model performance issues in production}
   \label{fig:identification_process}
\end{figure}

We asked participants about the processes they use to determine when an ML model is not functioning properly, i.e., model outcome problems. The response distribution is shown in Fig.~\ref{fig:identification_process}, with the most commonly mentioned method being user feedback, with 19 participants relying on real-world insights from end users. ML monitoring tools were the second most used method mentioned by 16 participants, reflecting the importance of automated solutions in large-scale deployments.

System crashes, chosen by 7 participants, were a less frequent method, as most respondents aim to detect issues before a system failure occurs. Additionally, 6 participants reported having no experience with ML in production, suggesting that some are still in the early stages of ML adoption.
Two participants selected ``Other" and specified that they use manual testing and model error analysis to identify problems with ML in production.

\begin{figure}[t]
  \centering
  \includegraphics[width=\linewidth]{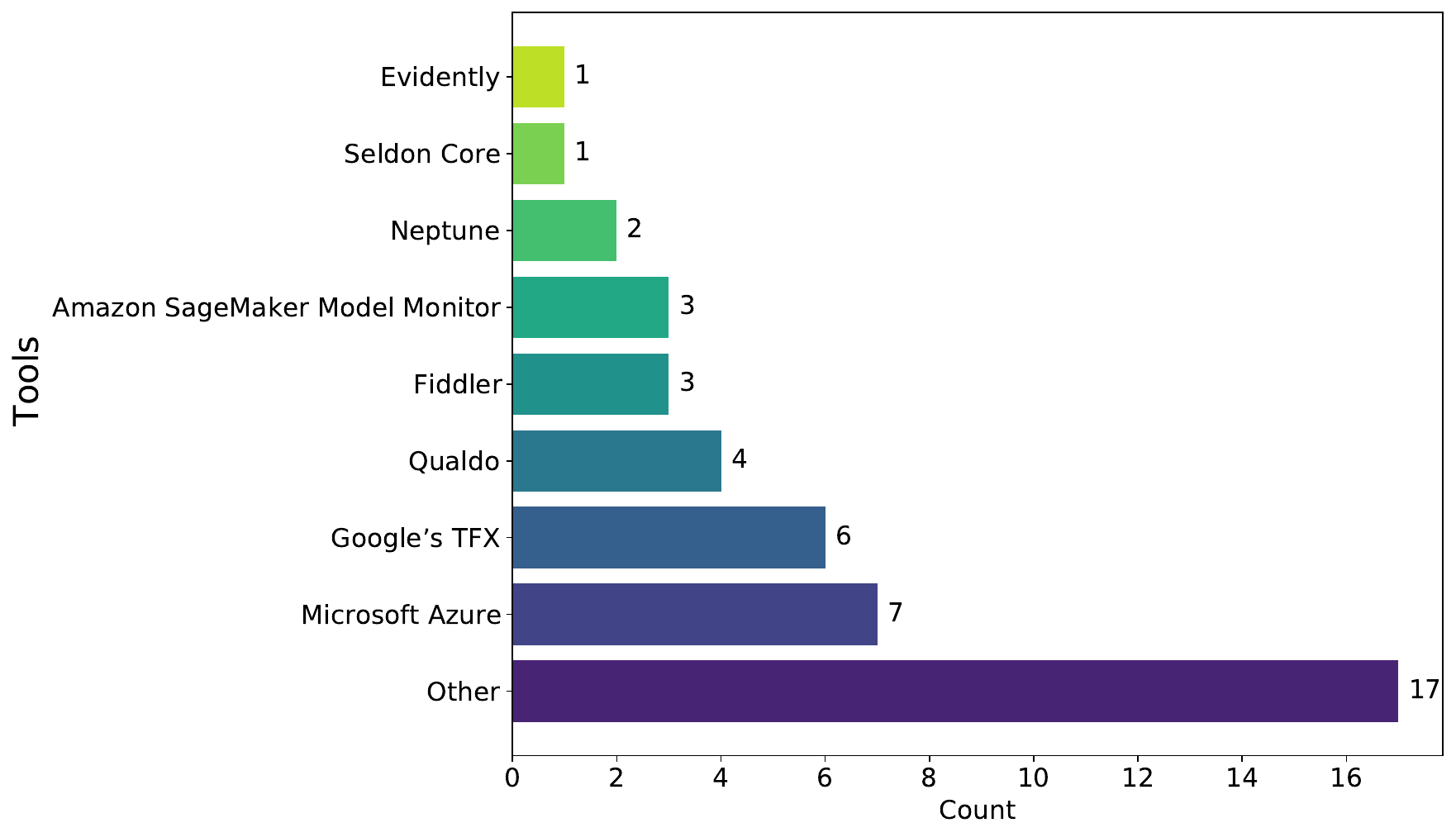}
  \caption{Tools used for monitoring production environments}
   \label{fig:monitoring_tools}
\end{figure}

\smallbreak
We asked participants about the tools they use to monitor production environments. The survey results shown in Fig.~\ref{fig:monitoring_tools} indicate that many participants utilise tools not listed among the predefined options. The most frequently mentioned category, ``Other" (17 responses), suggests that many participants employ monitoring solutions beyond the popular platforms.
Microsoft Azure was the second tool, with 7 responses, followed by Google’s TFX with 6 responses, and Qualdo with 4 responses. Tools such as Fiddler and Amazon SageMaker Model Monitor had moderate representation with 3 responses each. Meanwhile, Neptune, Seldon Core, and Evidently were less frequently used, with 2, 1, and 1 responses, respectively. This distribution highlights that while established tools like Azure and Google’s TFX are widely adopted, there is significant use of alternative or in-house monitoring solutions, as evidenced by the high number of ``Other" responses.

Next, we wanted to understand whether participants felt these tools met their expectations. While 17 participants expressed satisfaction, indicating that the tools generally meet their requirements, there were additional comments provided. Positive responses, such as ``Yes", ``Yes, to a certain extent", and ``It can basically meet our requirements", reflect that many users feel these tools achieve their intended purpose but are not without limitations. Some users mentioned that the tools sometimes fall short but are acceptable overall. For instance, one respondent noted that while the tools generally perform their intended functions, they do not fully accommodate the specialised needs of ML. Others highlighted that the tools are explainable, easy to understand, and customisable, especially when integrated with services like AWS for graph analysis. Some users acknowledged that while the tools may not be perfect, features like logging and application insights help. 

Negative responses reveal several challenges and limitations by participants. Key issues include:
\begin{itemize}
    \item Some participants indicated they have not used the tools before or are unfamiliar with any of the listed options, showing a reliance on manual debugging or self-developed solutions.
    \item Many participants mentioned they still perform manual debugging and have not integrated specific ML monitoring tools into their workflows. For these users, the tools either do not meet their needs or are not yet adopted.
    \item There were concerns about the difficulties of aligning the outputs of these tools with the actual behavior of their ML models and that these tools might be too biased.
    \item A common challenge mentioned was the need to use multiple tools to address diverse requirements. Some users found it difficult to find a single tool that meets all their needs, leading them to rely on a combination of tools and in-house solutions. This approach often requires significant manual effort and custom evaluation methods, highlighting the limitations of existing tools in fully addressing their specific ML needs. 
\end{itemize}

The feedback highlights gaps in addressing specialised ML requirements, and reducing manual efforts. It suggests a need for more comprehensive and integrated solutions that offer better flexibility and accuracy to meet diverse and complex needs in ML monitoring.

\begin{figure}[t]
  \centering
  \includegraphics[width=\linewidth]{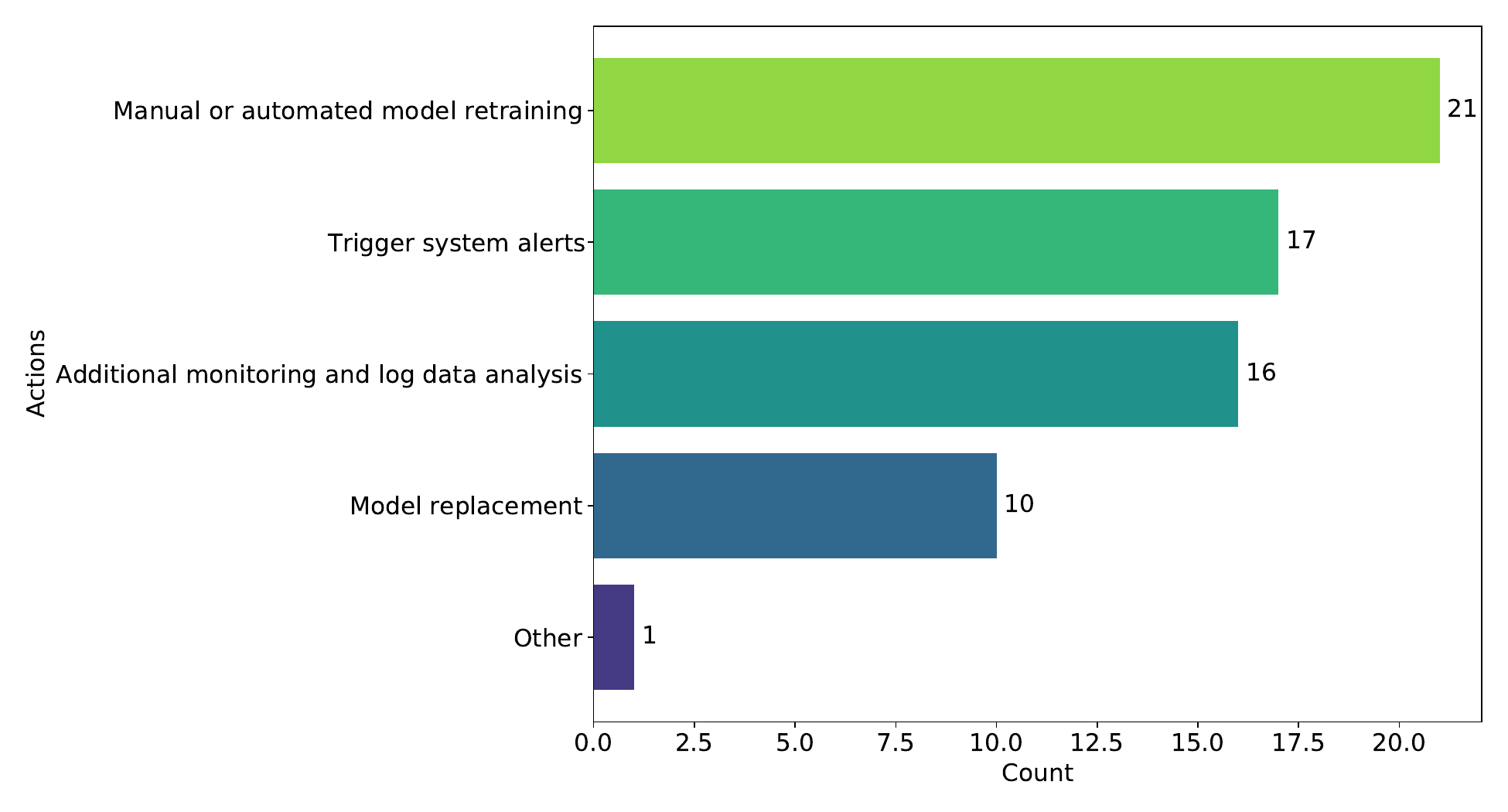}
  \caption{Recommended actions for ML model deviations in production}
   \label{fig:proper_actions}
\end{figure}
Finally, we asked participants about the appropriate actions when the production environment changes and ML models deviate from normal behavior. The distribution of responses is shown in Fig.~\ref{fig:proper_actions}. The most popular response is ``Manual or automated model retraining", with 21 responses. This suggests that many participants view retraining, either manually or automatically, as a crucial action to address deviations in model performance. ``Trigger system alerts" is the second most common response, with 17 responses. This indicates that many users prioritise being promptly informed about deviations, allowing for quicker responses and interventions. ``Additional monitoring and log data analysis" received 16 responses. This highlights the importance of ongoing monitoring and analysis in identifying deviations. It shows that users value detailed insights into model performance and system behavior to understand the nature and cause of the deviations. ``Model replacement" was chosen by 10 participants. While less common than retraining, alerts, or monitoring, some users see replacing the model as a viable solution when deviations happen. One participant selected ``Other" and suggested that the proper action is to ``revert the change if possible". This response highlights a more conservative approach, where instead of focusing on retraining or replacing the model, the participant prefers to undo the recent changes that may have caused the deviation. This would allow the system to return to a stable state without further adjustments to the model itself.

\smallbreak
\noindent\textbf{Protocol Design:}
\label{sec:protocol_design}

The survey responses offer valuable insights into the key aspects to consider when designing a wrapper around ML components in production environments. 
Fig.~\ref{fig:wrapper_components}
presents the distribution of responses, with a detailed analysis of the feedback provided below:

\begin{enumerate}
    \item \textbf{Input validation:} A total of 43.3\% of participants either agree or strongly agree that input validation is essential, such as verifying image dimensionality and contrast in computer vision tasks. However, a notable 56.7\% strongly disagree. The divided response suggests the importance of input validation may change depending on the specific use case, particularly in scenarios where data is already pre-validated or models include inherent validation mechanisms.

    \item \textbf{Security and adversarial attack detection:} The majority of participants (76.7\%) agree or strongly agree that evaluating inputs for security violations is crucial for ensuring the robustness of ML-enabled software systems. In contrast, only 2 participants disagree, while 5 participants either have no opinion or are neutral. There is broad agreement on the importance of safeguarding ML models against adversarial attacks, highlighting the critical role of security in maintaining the robustness and reliability of production systems.
\begin{figure}[t]
  \centering
  \includegraphics[width=\linewidth]{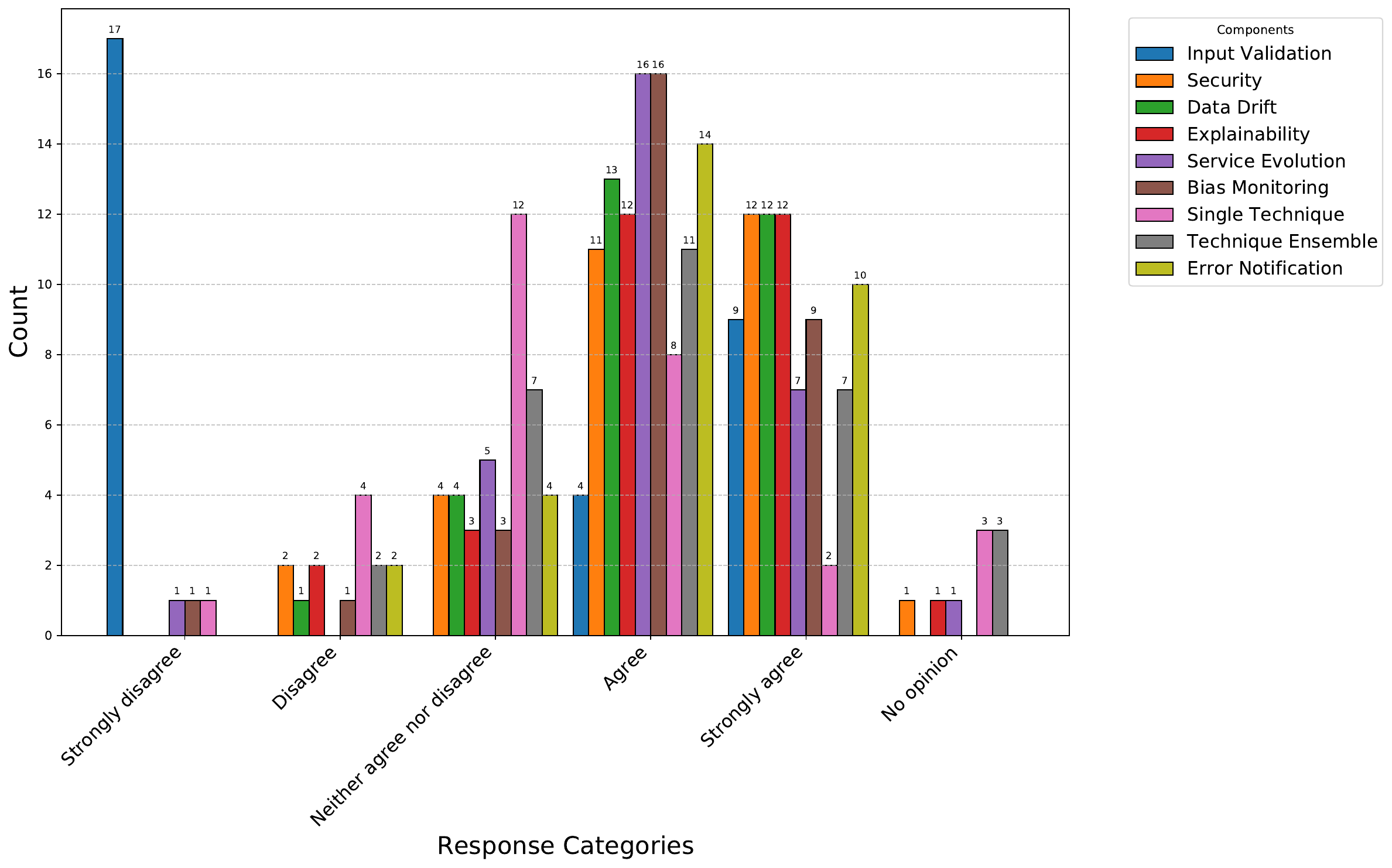}
  \caption{Key considerations when designing a wrapper around ML components in production}
   \label{fig:wrapper_components}
\end{figure}

    \item \textbf{Data drift detection:} A significant majority (83.3\%) of participants agree or strongly agree that monitoring the evolution of production environments to detect data drift is crucial for ensuring model robustness over time. Data drift monitoring is widely acknowledged as a key component in designing a wrapper around ML components in production to ensure the robustness of the systems employing them.

    \item \textbf{Explainability:} A total of 24 participants (80\%) consider it important to provide explanations for the decisions made by ML models. In contrast, only 2 participants disagreed, while 5 either neutral or had no opinion. The need for transparency and explainability in ML systems is highly valued to enhance robustness and trust in production environments.

    \item \textbf{Service evolution monitoring:} A total of 23 participants (76.7\%) agree or strongly agree that monitoring changes in ML services over time, such as shifts in labels and confidence scores, is critical for ensuring continuous performance. Only 1 participant strongly disagrees, and 6 are neutral or have no opinion. Monitoring the evolution of ML services is essential for maintaining the robustness of systems, particularly in evolving production environments.

    \item \textbf{Bias monitoring and fairness:} A total of 25 responses (83.3\%) agree or strongly agree that monitoring ML services for bias and ensuring fairness is essential. In contrast, 2 participants disagree or strongly disagree, and 3 are neutral. Ethical considerations, such as bias reduction and ensuring fairness, are crucial for robust and responsible ML-enabled software systems in production

    \item \textbf{Single technique approach:} Only 10 participants (33.3\%) agree or strongly agree that using a single technique is sufficient to address production issues, while 15 participants (50\%) are neutral and 5 participants disagree or strongly disagree. Most participants prefer more comprehensive approaches instead of a uniform solution for addressing ML production challenges.

    \item \textbf{Techniques ensemble:} A majority of 60\% of participants favor using a combination of techniques and ensemble methods to address production ML problems, while 2 disagree, and 10 are neutral or have no opinion. There is a preference for using ensemble of methods as an effective approach to handling complex issues in production. Leveraging multiple techniques can improve robustness and performance. For example, using MSP and energy score to detect data drift is preferred over relying on MSP as a single technique for detection.

    \item \textbf{Error notification and codes:} A majority of 24 participants (80\%) agree or strongly agree that having a well-defined notification mechanism and a set of error codes to describe changes in production is essential. In contrast, only 2 participants disagree and 4 are neutral. A structured error notification and reporting system is viewed as a critical component for ensuring timely detection and communication of issues related to ML models in production.

\end{enumerate}

The survey responses reveal the importance of data drift detection, security, and explainability in ML-enabled software systems. A preference for ensemble techniques over single methods indicates the need for more comprehensive approaches to handling production issues. Additionally, there is support for structured error notification systems and bias monitoring. Overall, these findings highlight the need for robust, flexible approaches to maintain and ensure the robustness and trustworthiness of ML models in production.

\smallbreak
\noindent\textbf{Protocol Design Feedback:}

In this part of the survey, we aimed to gather additional insights from participants on other important aspects of an ML wrapper from their perspective, and the responses revealed the following:
\begin{itemize}
    \item Sixteen participants either had no suggestions or believed that the discussed aspects were adequate. This may indicate a preference for focusing on existing features rather than introducing new ones.
    \item Participants suggest that an ML wrapper should include functionalities for comparing model performance via A/B testing, defining output boundaries to ensure robustness, and tracking software versions and conflicts.
    \item Suggestions for continual learning and enabling user feedback to drive model adaptation, either manually or automatically. 
    \item Several responses highlight the importance of logging, tracking additional attributes (like response time and traffic), and comprehensive documentation. These elements are vital for troubleshooting, reproducibility, and understanding model behavior.
    \item Participants also highlighted the need for interoperability with popular ML frameworks and libraries and the importance of API simplicity and ease of use. This suggests that the wrapper should seamlessly integrate with existing tools and be user-friendly.
    \item Participants emphasised that the ML wrapper should include specific coverage for challenges related to LLMs, indicating the need to address the unique issues associated with these models.
\end{itemize}

\smallbreak
\noindent\textbf{ML-On-Rails Protocol Evaluation:}

We have implemented the ML-On-Rails protocol to address critical ML problems encountered in production, including: input validation, OOD detection, adversarial attack detection, model explainability and error codes.

According to the previous results in Section~\ref{sec:protocol_design}, OOD detection, adversarial attack detection, model explainability, and error codes are considered critical. However, some participants perceive input validation as less critical. To evaluate how well the proposed protocol addresses these issues, we asked participants to assess the protocol components using the example shown in Fig.~\ref{fig:user_survey_protocol}. The distribution of participants' responses regarding the proposed protocol components is illustrated in Fig.~\ref{fig:protocol_guards}, where: 

\begin{figure}[t]
  \centering
  \includegraphics[width=\linewidth]{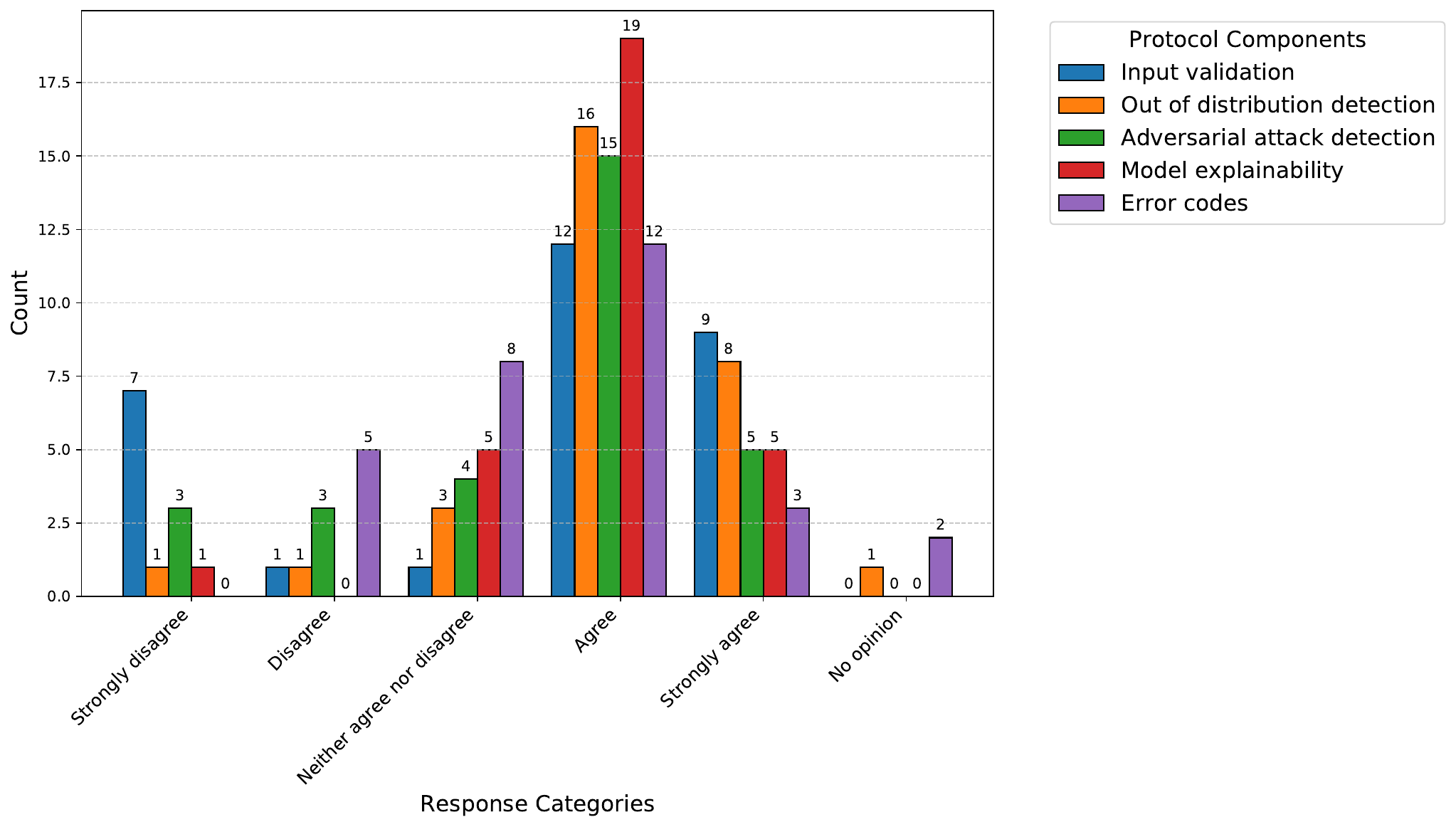}
  \caption{Effectiveness of key protocol aspects for ML production issues}
   \label{fig:protocol_guards}
\end{figure}

\begin{enumerate}
    \item \textbf{Input validation:} 21 participants (70\%) agree or strongly agree that input validation is effective, increased from the 43.3\% in earlier feedback. Eight participants disagree or strongly disagree, and 1 is neutral. This change suggests that seeing the input validation guard in action influenced participants' views positively.

    \item \textbf{OOD detection:} 24 participants (80\%) agree or strongly agree, consistent with earlier feedback. Two participants disagree, and 4 are neutral or have no opinion. This consistency highlights OOD detection as a critical component.

    \item \textbf{Adversarial attack detection:} 20 participants agree or strongly agree, 4 are neutral, and 6 disagree or strongly disagree. This indicates some variation in opinions, though it remains a significant area of concern.

    \item \textbf{Model explainability:} 24 participants (80\%) agree or strongly agree, with 1 strongly disagreeing and 5 being neutral. This shows strong support for the importance of model explainability.

    \item \textbf{Error codes:} 15 participants (50\%) agree or strongly agree, signifying a 30\% reduction from previous responses. Meanwhile, 10 participants are either neutral or have no opinion, and 5 participants disagree. A notable proportion of participants are neutral or indifferent, which differs from previous responses.
\end{enumerate}

Overall, the feedback indicates that while most participants find the proposed protocol components effective, there is a change in the perceived importance of each aspect, with particular emphasis on OOD detection, adversarial attack detection, and model explainability.

\begin{figure}[t]
  \centering
  \includegraphics[width=\linewidth]{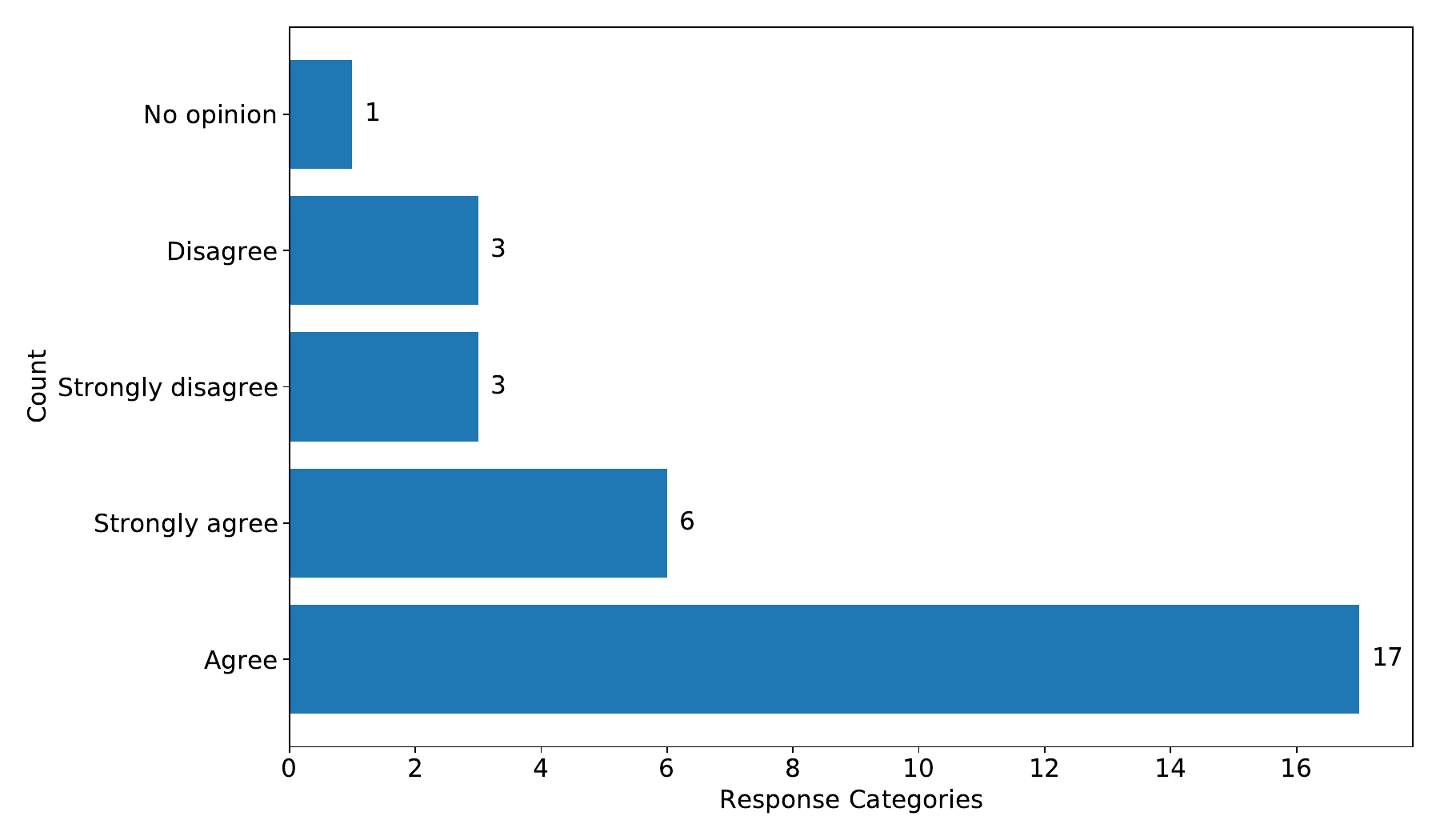}
  \caption{Participants opinions on ML protocol usability}
   \label{fig:ease_of_use}
\end{figure}

\smallbreak
When we asked the participants about the protocol's ease of use, 76.7\% agreed or strongly agreed that the proposed ML protocol is easy to understand and use, indicating generally positive feedback. However, 20\% of the participants disagreed or strongly disagreed, suggesting that some users found the protocol challenging. This feedback as illustrated in Fig.~\ref{fig:ease_of_use} highlights the need for potential improvements in the protocol's clarity and usability. 

\begin{figure}[t]
  \includegraphics[width=\linewidth]{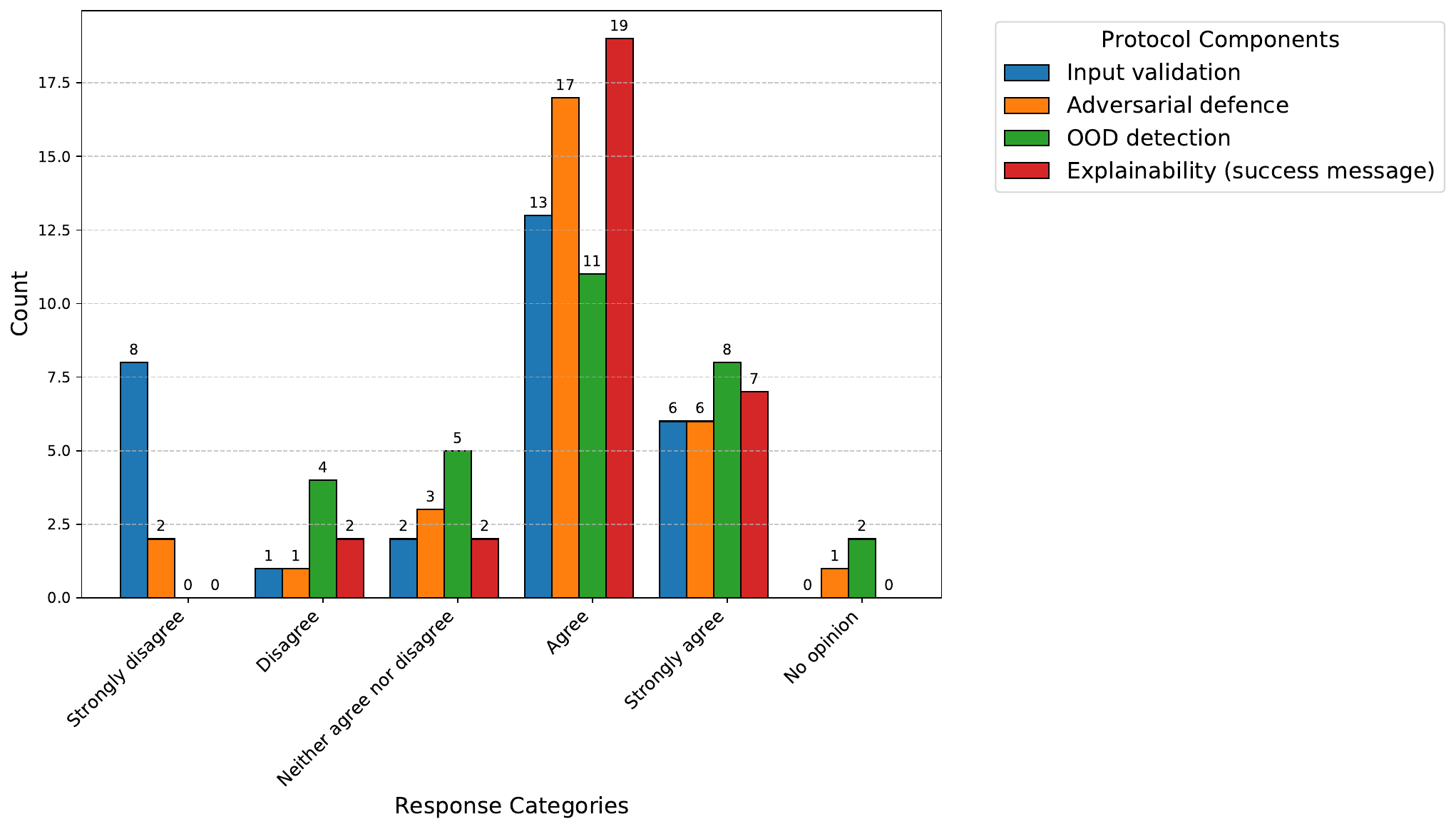}
  \caption{Participants responses on the clarity and informativeness of protocol components messages}
   \label{fig:error_message}
\end{figure}

\smallbreak
The evaluation of the clarity and informativeness of the error/success messages across various protocol components had insightful results, as illustrated in Fig.~\ref{fig:error_message}.

\noindent\textbf{Input validation:} A majority of participants (63.3\%) agreed or strongly agreed that the error messages were clear and informative, with 9 participants (30\%) disagreeing or strongly disagreeing. This indicates that while many found the messages helpful, some participants had concerns regarding the clarity of these messages.

\noindent\textbf{Adversarial defense:} This component received positive feedback, with 76.7\% of participants agreeing or strongly agreeing that the messages were clear and informative. Only 10\% disagreed, and 4 responses were neutral. This indicates general satisfaction with the messages for adversarial defenses.

\noindent\textbf{OOD detection:} 
63.3\% of participants agreed or strongly agreed that the OOD detection messages were clear, while 13.3\% disagreed and 23.3\% remained neutral or had no opinion. This shows moderate support but also highlights that the clarity of OOD messages could be improved.

\noindent\textbf{Explainability and success message:} The highest positive feedback came from explainability messages, with 86.7\% of participants agreeing or strongly agreeing that these were clear and informative. Only 6.6\% were neutral, and 6.6\% disagreed, demonstrating strong approval for this component.

While the overall feedback on the clarity and informativeness of the messages is positive, particularly for adversarial defense and explainability, input validation and OOD detection could be improved to ensure clarity for more users.

\smallbreak
The final question in the survey asked participants for their overall thoughts on the proposed ML protocol and whether they believed any aspects were missing, needed improvement, or required changes. The feedback provided by participants highlights several key areas for improvement, offering valuable insights for enhancing the protocol's robustness and usability:
\begin{itemize}
    \item \textbf{Error message clarity and specificity:}
    Several participants mentioned that the error messages during input validation are too generic and could be more informative. Specific feedback pointed out that having one general error message for multiple conditions might reduce the usefulness of the message. Participants suggested that different error codes and more detailed messages would be helpful, such as clarifying how to resolve validation issues or specifying which part of the input was manipulated in cases of adversarial attacks.
        A suggested improvement is to use specific error codes, such as 4xx for client-side errors (e.g., input validation) and 5xx for server-side issues would make debugging easier and more intuitive.
    \item \textbf{Security and adversarial defense:} Some participants expressed concern about the transparency of the adversarial defense response. They suggested that openly flagging an input as adversarial could provide malicious attackers with information to reverse-engineer the defense mechanisms. Instead, it was recommended to return a generic error while logging the specific error internally to avoid exposing the system’s defense strategies. Additionally, participants requested more detailed information about detected adversarial attacks, such as specifying where the attack was applied (e.g., localisation in computer vision models) or providing the severity level of the attack.

    \item \textbf{Guards order of execution:} One participant questioned the order of operations in the protocol, suggesting that OOD detection should be performed before passing inputs to the model.
    \item \textbf{Latency evaluation:} One participant suggested to include latency evaluation, and methods to validate model outputs. 
    \item \textbf{Implementation complexity:} One participant acknowledged that the protocol effectively addresses critical issues in production ML but expressed concerns about the effort required to implement it for each model.

    \item \textbf{Positive feedback:} Overall, participants recognised the protocol as a positive step toward making ML models more secure, robust, and predictable. Several participants mentioned that the protocol effectively addresses critical ML production challenges and that it provides essential information for both developers and end-users. The simplicity and effectiveness of the approach were appreciated, with participants noting that it would work well in systems that log errors or track issues.

\end{itemize}

\section{Discussion}
\label{sec:discussion}

This section summarises the survey results, focusing on the challenges software engineers face when working with ML models in production, identifying issues, and managing these models effectively. The findings provide valuable insights into the complexities of model deployment and highlight the areas requiring improvements. Based on participant feedback, the ML protocol will be refined to enhance its effectiveness and address the challenges identified in the survey.

\subsection{Challenges Software Engineers Face with ML Models}

The survey results highlighted several key challenges when working with machine learning models in understanding specific ML jargon and integrating these models into production environments. Terms such as ``Model uncertainty", ``Out-of-distribution data", and ``Online prediction" emerged as the most difficult to comprehend and communicate, reflecting the complexity of these concepts for many professionals.

Integrating ML models into software systems comes with significant challenges, with participants identifying the lack of explainability, issues with model versioning, and unrealistic expectations as the top concerns. 
Furthermore, insufficient documentation on models, training processes, and datasets was a recurring theme, indicating a need for more transparent and comprehensive resources.

In production environments, the most critical challenges reported were silent failures, out-of-distribution data, and lack of transparency. Unpredictable outputs, model uncertainty, and concerns related to adversarial attacks and fairness monitoring were also major issues. Overall, these findings emphasise the necessity of better tools, improved documentation, and enhanced monitoring to ensure the robustness, transparency, and reliability of ML models in practice.

\subsection{Identifying ML Model Issues in Production}
To identify issues with ML models in production, the survey results highlighted several key challenges in monitoring and managing these models. User feedback and ML monitoring tools emerged as the most common methods for detecting when a model is not functioning properly, reflecting the importance of both real-world insights and automated systems.

The tools used to monitor production environments highlight that many participants rely on alternative or custom solutions beyond popular platforms like Microsoft Azure and Google’s TFX. This reliance on ``Other" tools highlights gaps in existing monitoring solutions, indicating the need for more comprehensive, flexible, and specialised tools to better meet the complex demands of ML in production. 

While some participants are satisfied with their current tools, many highlighted limitations, like the need for manual effort, difficulties aligning tool outputs with model behavior, and the necessity of using multiple tools to cover all requirements. These insights point to the need for solutions that can reduce manual intervention, enhance transparency, and improve the overall robustness of ML models in production environments.

Finally, the survey responses indicate that participants consider different strategies when ML models deviate from expected behavior in production environments. The most common approach is manual or automated retraining, highlighting the importance of regularly updating models to address performance issues. Triggering system alerts and conducting additional monitoring and log data analysis were also highly valued, demonstrating the need for timely detection and detailed insight into model deviations. 

\subsection{Protocol Design}
The survey results provided valuable insights into the key considerations when designing a wrapper around ML components for production environments. The responses emphasised the critical importance of several components: data drift detection, security (including adversarial attack detection), and explainability. The preference for ensemble techniques over single ones suggests that comprehensive strategies are needed to address the complex issues in production environments. Participants also highlighted the significance of structured error notification systems and monitoring for bias and fairness, both essential for building trustworthy and responsible ML systems. Additionally, participants suggested addressing challenges related to LLMs, reflecting the growing complexity of modern ML-enabled software systems.

\subsection{ML Protocol Evaluation}
The participants evaluation of the proposed ML-On-Rails protocol revealed its effectiveness in addressing critical ML issues in production environments, particularly OOD detection, adversarial attack detection, and model explainability. These components were highly rated, highlighting their importance in enhancing the robustness of ML-enabled systems.

Input validation, initially perceived as less critical, received increased support after participants observed its practical implementation. However, feedback on error messages highlighted areas for improvement, particularly in the clarity and specificity of input validation and OOD detection messages.

The usability of the protocol received  positive feedback, with 76.7\% of participants agreeing that the protocol was easy to use. Participants offered valuable insights for enhancing the protocol, such as providing more specific error messages, optimising guard execution order, and considering the inclusion of latency evaluations. Additionally, participants recommended avoiding the exposure of detailed feedback on adversarial defense responses that could help potential attackers.

\subsection{Key Improvements Based on Participants Feedback}

We analysed the participants' responses to refine our proposed protocol based on their feedback. The key areas of improvement include:
\begin{itemize}
    \item \textbf{Error codes:} Enhanced debugging with more specific HTTP status codes. Client-side issues return 4xx codes, like HTTP 400 for input validation failures, while server-side failures, such as OOD data detection, trigger 5xx codes. Each error provides detailed messages for quick diagnosis and resolution.
    \item \textbf{LLM challenges:} Focused on addressing LLM challenges to improve the robustness of ML-enabled systems.
\end{itemize}


In addition to these improvements, several clarifications based on participant feedback are essential for a comprehensive understanding of our protocol:
\begin{itemize}
    \item \textbf{Guard order of execution:} OOD detection should ideally be performed before inputs are passed to the model to ensure that invalid data is filtered out early. However, the appropriate point of detection depends on the technique being used. In our proposed protocol, techniques such as MSP and max-logit require model outputs to detect OOD data. Assuming access to training data, we could use the Hellinger distance~\cite{abdelkader2020towards} as an OOD detector to identify OOD inputs before inference.
    \item \textbf{Security and adversarial defense:} Participants expressed concerns that adversarial defense alerts could provide attackers with insights to reverse-engineer the defense. However, these alerts are intended for engineers, not end-users, to help them understand the error and take appropriate actions without exposing sensitive details.
\end{itemize}

\section{Conclusion}
\label{sec:conclusion}
Most participants acknowledged that the proposed ML-On-Rails is valuable in enhancing robustness and providing actionable insights. Positive feedback was received for OOD detection, adversarial defense and explainability, though some reported that input validation could be clearer and more refined. The protocol’s simplicity and ease of implementation were highly valued. Based on the feedback, we will refine the protocol to better support software engineers. We will improve the debugging process by utilising more specific HTTP status codes. Also, we will integrate safeguards specifically designed for LLMs.

\newpage
\bibliographystyle{IEEEtran}
\bibliography{references}

\end{document}